\documentclass[twocolumn,showpacs,preprintnumbers,amsmath,amssymb,superscriptaddress,showkeys,prd]{revtex4}

\usepackage{graphicx}
\usepackage{dcolumn}
\usepackage{bm}
\usepackage{amsmath,amssymb}

\begin{document}

\preprint{\emph{Revista Mexicana de Astronom\'{\i}a y
Astrof\'{\i}sica}}

\title{Spectra of HB 21 supernova remnant: Evidence of spectra flattening at the low frequencies}

\author{D. Borka}
\email[Corresponding author:]{dusborka@vinca.rs} \affiliation{Atomic
Physics Laboratory (040), Vin\v{c}a Institute of Nuclear Sciences,
University of Belgrade, P.O. Box 522, 11001 Belgrade, Serbia}

\author{V. Borka Jovanovi\'{c}}
\affiliation{Atomic Physics Laboratory (040), Vin\v{c}a Institute of
Nuclear Sciences, University of Belgrade, P.O. Box 522, 11001
Belgrade, Serbia}

\author{D. Uro\v{s}evi\'{c}}
\affiliation{Department of Astronomy, Faculty of Mathematics,
University of Belgrade, Studentski trg 16, 11000 Belgrade, Serbia}

\date{November 7, 2011}

\begin{abstract}
We use observations of the continuum radio emission at 1420, 820,
408, 34.5 and 22 MHz to estimate the mean brightness temperatures of
the HB 21 supernova remnant (SNR) at five frequencies. We also
presented mean spectral index of HB 21. The spectra of HB 21 are
estimated for mean temperatures versus frequency for 1420, 820, 408,
34.5 and 22 MHz. We also presented $T-T$ plots of three frequency
pairs: between 1420--34.5, 1420--22, 34.5--22 MHz. We noticed
flatter spectral indices at frequencies below 408 MHz. Probably this
is due to the absorption by thermal plasma at low frequencies.
\end{abstract}


\maketitle

\section{Introduction}
\label{sec01}

The radio emission from supernova remnants (SNRs) is generally
understood to be synchrotron emission from relativistic electrons
moving in magnetic fields.

The HB 21 remnant is listed in Green's catalogue of Galactic
supernova remnants \citep{gree09a,gree09b} as G89.0+4.7. It has
mixed morphology, e.g., shell-like in radio and center-filled in the
X-ray \citep{rho86}. As described in \citet{koth06}, this is an old
SNR evolving through the later stages of the Sedov phase, or early
into the radiative (where a significant amount of the shock energy
is being radiated away). Results from \citet{byun06} suggest that
the unusual radio features and the central thermal X-ray
enhancements of HB 21 might be the result of an interaction with
molecular clouds. One form of evidences for the interaction between
the SNR and the molecular cloud in HB 21 is the discovery of shocked
molecular clumps by \citet{koo01}. The center-filled, thermal X-ray
emission is suggested to be caused by interaction with molecular
clouds.

Flux-calibrated images and spectra of the extended remnant HB 21 in
the optical range, are presented in \citet{mavr07}. Filamentary and
patchy structures were detected and appear to be correlated with the
radio emission.

Significant spatial variations in spectral index are found in HB 21
\citep{leah06}. He consider different physical mechanisms for
spectral index changes, including a detailed consideration of
ionization losses in the dense molecular gas interacting with HB 21.

One of the main goals of this paper is to provide evidence for
Leahy's \citep{leah06} theoretical proposal. He suggests that if the
spectra below 408 MHz show flatter spectral indices then absorption
by thermal plasma would be the preferred mechanism. We show the
evidence of spectrum flattening at the low radio frequencies in the
supernova remnant HB 21.

\begin{table*}
\centering
\caption{Temperatures and brightnesses of HB 21 at 1420,
820, 408, 34.5 and 22 MHz.}
\begin{tabular}{c|c|c|c}
\hline
Frequency & Temperature limits & Temperature & Brightness \\
(MHz) & $T_\mathrm{min}$, $T_\mathrm{max}$ (K) & (K) & (10$^{-22}$ W/(m$^2$ Hz Sr)) \\
\hline
1420 & 6.0, 9.0 & 1.50 $\pm$ 0.05 & 9.26 $\pm $ 0.30 \\
820 & 16.5, 26.0 & 4.91 $\pm$ 0.20 & 10.13 $\pm$ 0.40 \\
408 & 87, 140 & 32.2 $\pm$ 1.0 & 16.43 $\pm$ 0.50 \\
34.5 & 49000, 81000 & 18237 $\pm$ 700 & 66.60 $\pm$ 2.56 \\
22 & 105000, 144000 & 28530 $\pm$ 500 & 43.34 $\pm $ 0.76 \\
\hline
\end{tabular}
\label{tab01}
\end{table*}

\begin{figure}[ht!]
\centering
\includegraphics[width=0.45\textwidth]{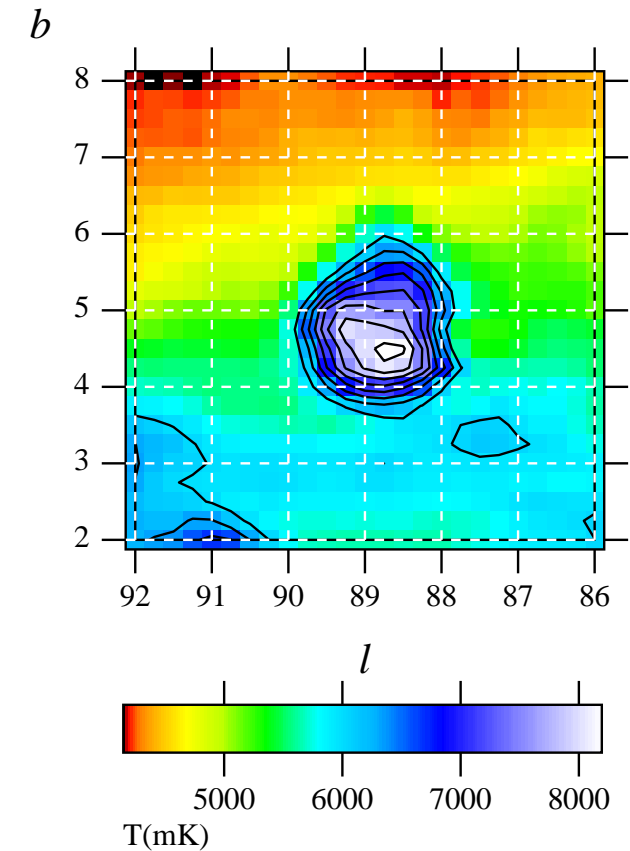} \\
\vspace*{0.8cm}
\includegraphics[width=0.48\textwidth]{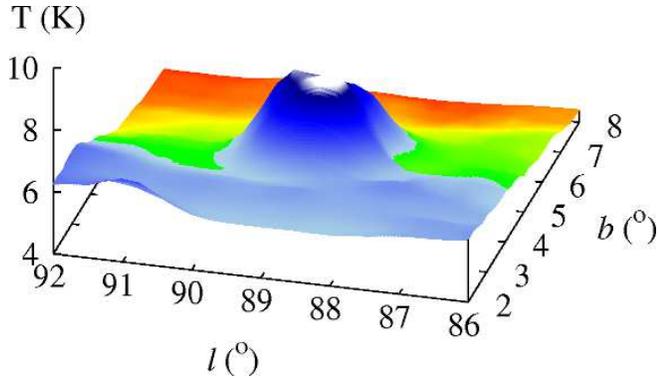} \\
\caption{\emph{Top}: The radio map of HB 21 SNR at 1420 MHz with
contours of brightness temperature. This SNR has position: $l$ =
[90.$^\circ$, 87.5$^\circ$]; $b$ = [3.5$^\circ$, 6$^\circ$], in new
Galactic coordinates ($l$, $b$). Eleven contours plotted represent
the temperatures $T_\mathrm{min}$ and $T_\mathrm{max}$ from Table
\ref{tab01} and nine contours in between. The contour interval is
0.3 K $T_\mathrm{b}$, starting from the lowest temperature of 6 K up
to 9 K. The corresponding temperature scale is given. \emph{Bottom}:
the 1420 MHz area map of Cygnus.}
\label{fig01}
\end{figure}

\begin{figure}[ht!]
\centering
\includegraphics[width=0.45\textwidth]{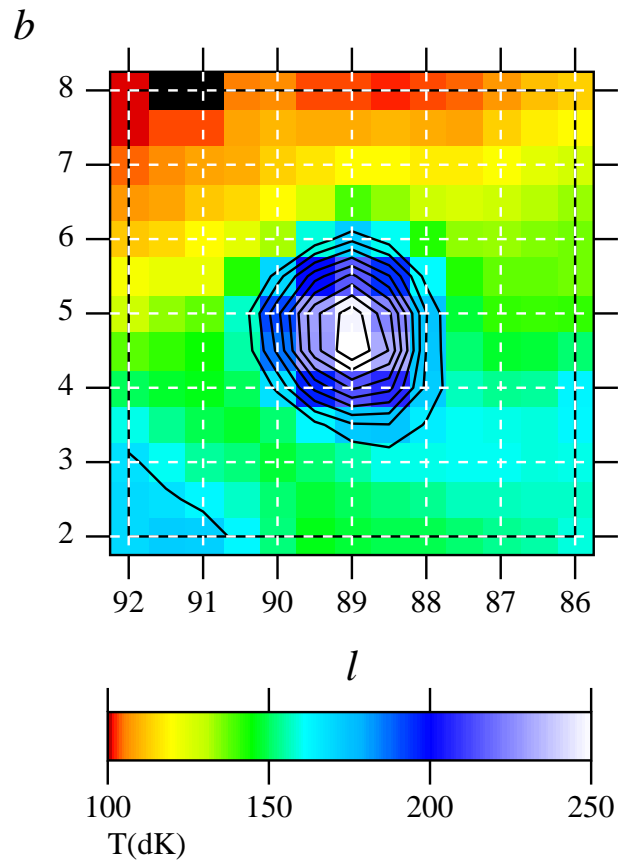} \\
\vspace*{0.8cm}
\includegraphics[width=0.48\textwidth]{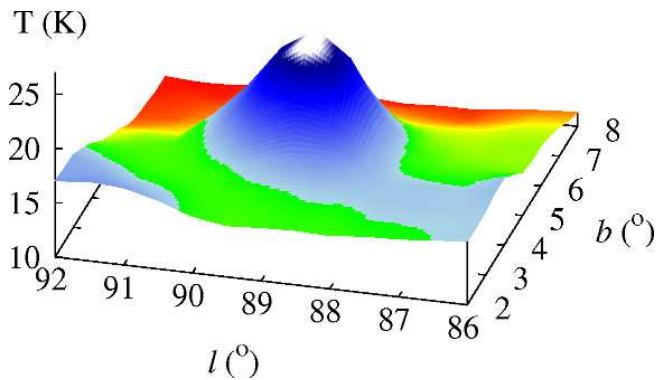} \\
\caption{The same as Fig. 1, but for 820 MHz. The contour interval
is 9.5 K $T_\mathrm{b}$, starting from the lowest temperature of
16.5 K up to 26 K.}
\label{fig02}
\end{figure}

\begin{figure}[ht!]
\centering
\includegraphics[width=0.45\textwidth]{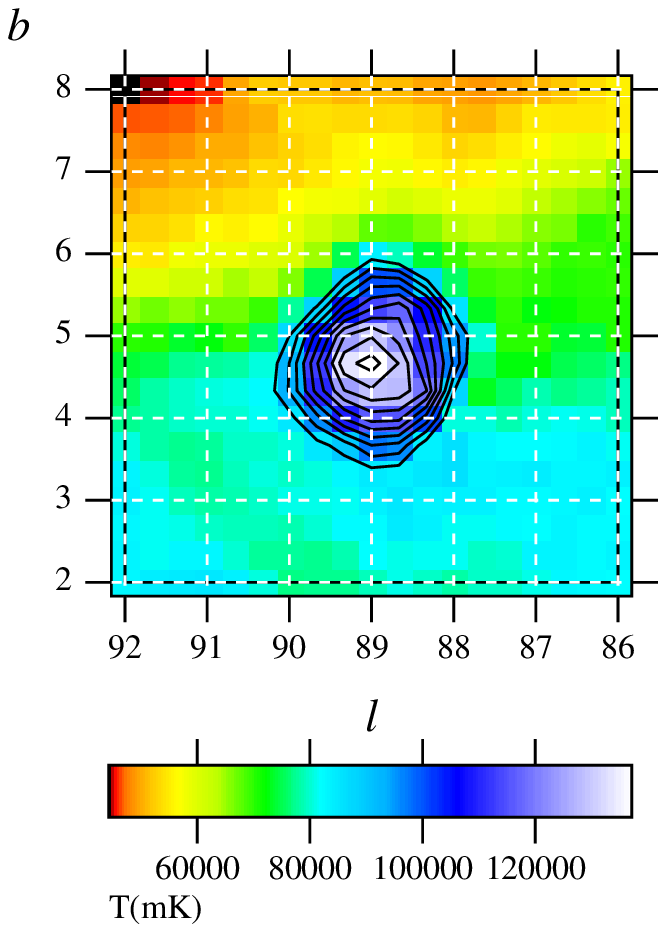} \\
\vspace*{0.8cm}
\includegraphics[width=0.48\textwidth]{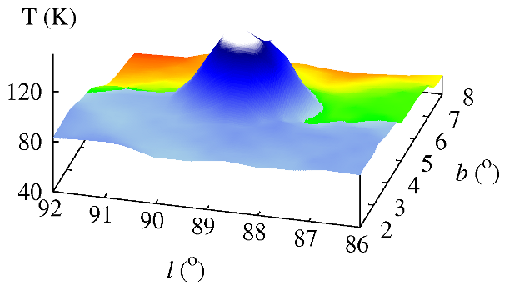} \\
\caption{The same as Fig. 1, but for 408 MHz. The contour interval
is 5.3 K $T_\mathrm{b}$, starting from the lowest temperature of
87 K up to 140 K.}
\label{fig03}
\end{figure}

\begin{figure}[ht!]
\centering
\includegraphics[width=0.45\textwidth]{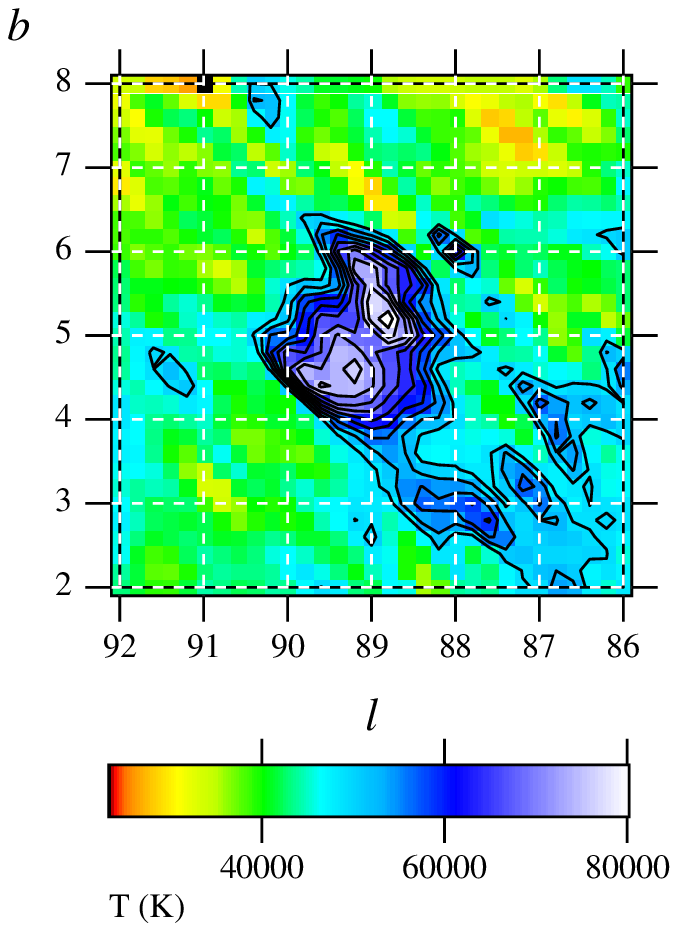} \\
\vspace*{0.8cm}
\includegraphics[width=0.48\textwidth]{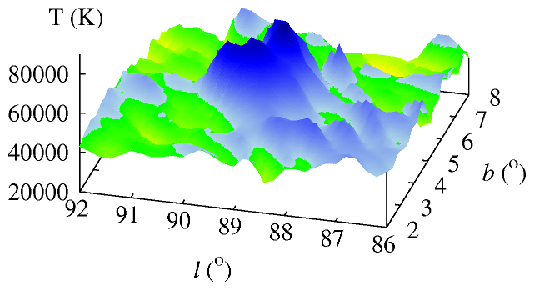} \\
\caption{The same as Fig. 1, but for 34.5 MHz. The contour interval
is 3200 K $T_\mathrm{b}$, starting from the lowest temperature of
49000 K up to 81000 K.}
\label{fig04}
\end{figure}

\begin{figure}[ht!]
\centering
\includegraphics[width=0.45\textwidth]{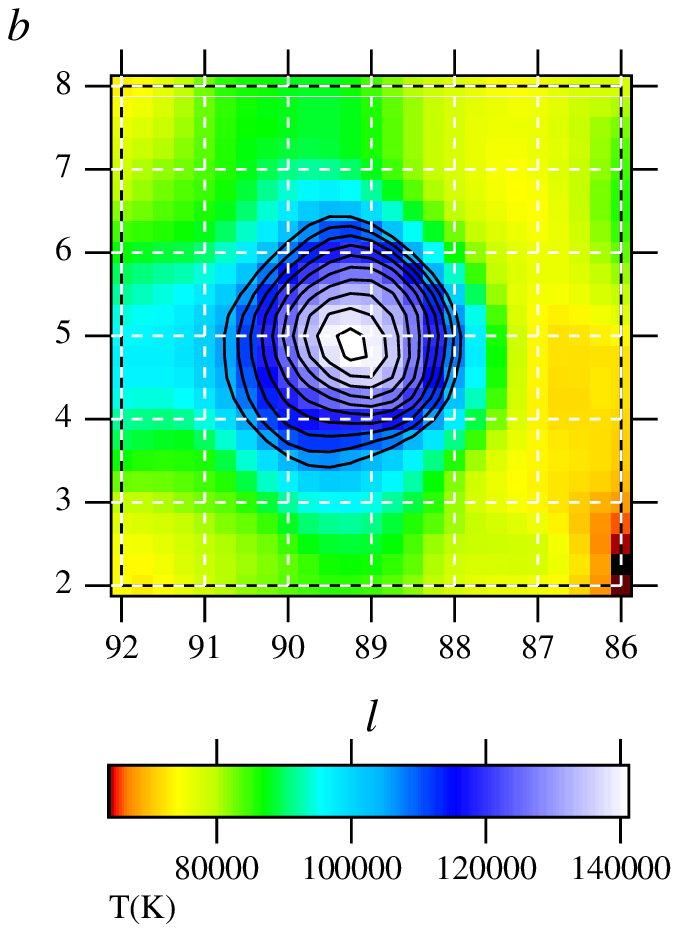} \\
\vspace*{0.8cm}
\includegraphics[width=0.48\textwidth]{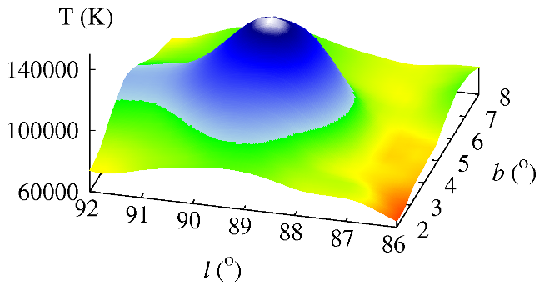} \\
\caption{The same as Fig. 1, but for 22 MHz. The contour interval
is 3900 K $T_\mathrm{b}$, starting from the lowest temperature of
105000 K up to 144000 K.}
\label{fig05}
\end{figure}

\begin{figure}[ht!]
\centering
\includegraphics[width=0.45\textwidth]{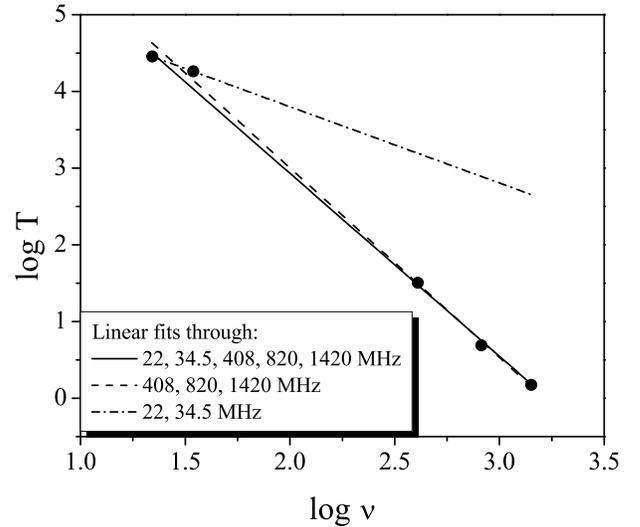}
\caption{The radio continum spectrum of HB 21 SNR: temperature
versus frequency. Spectrum for five measurements -- at 22, 34.5,
408, 820 and 1420 MHz is given by solid line, spectrum obtained from
the three frequencies -- 408, 820 and 1420 MHz is given by dashed
line, and spectrum obtained from the two lowest frequencies -- 22
and 34.5 MHz is given by dash-dot line.}
\label{fig06}
\end{figure}

\begin{figure}[ht!]
\centering
\includegraphics[width=0.45\textwidth]{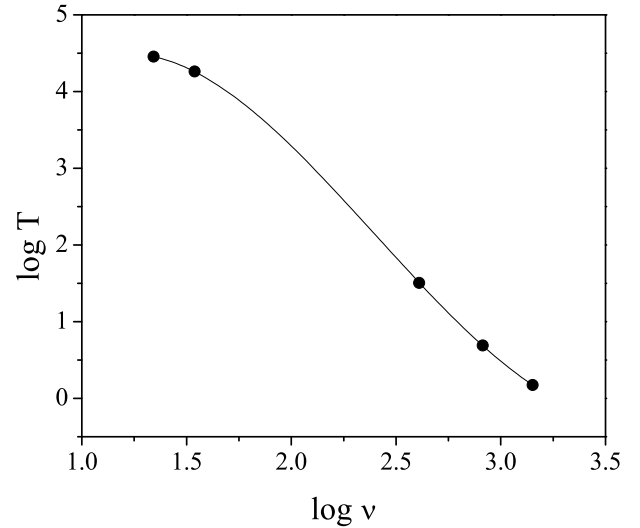}
\caption{The curved spectrum of the third order of HB 21,
obtained from the five frequencies.}
\label{fig07}
\end{figure}

\begin{figure}[ht!]
\centering
\includegraphics[width=0.45\textwidth]{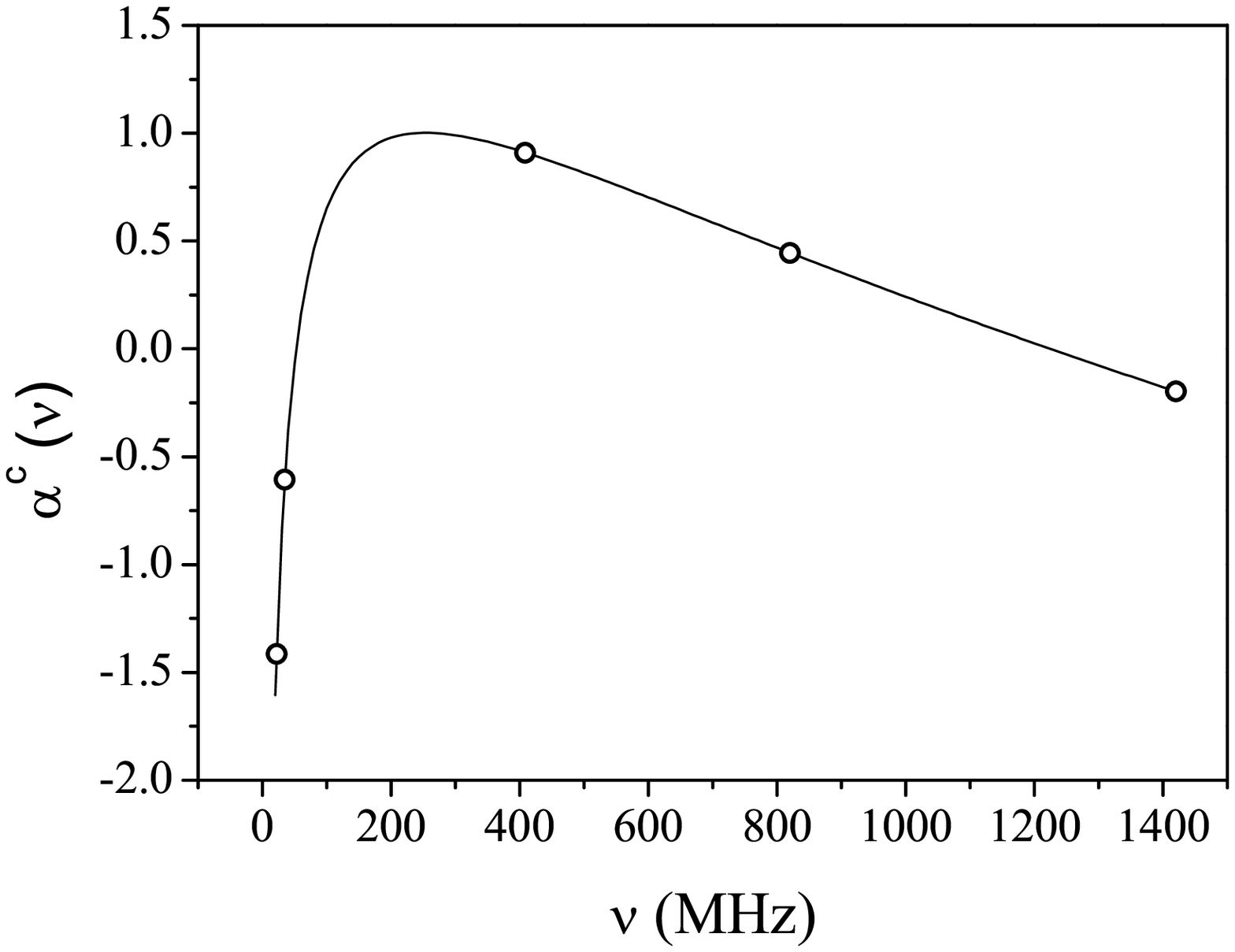}
\caption{Variations of spectral index $\alpha^c(\nu)$ with frequency in the case of
HB 21. Spectral index $\alpha^c(\nu)$ is obtained from the third order polynomial
fit presented in Fig. 7. Open circles represent values of spectral
indices $\alpha^c(\nu)$ at the five observed frequencies.}
\label{fig08}
\end{figure}

\section{Data and Method}

Measured data at five frequencies are used: 1420 \citep{reic86}, 820
\citep{berk72}, 408 \citep{hasl82}, 34.5 \citep{dwar90} and 22 MHz
\citep{roge99}. These surveys are available in electronic form in
"Flexible Image Transport System" (FITS) data format, at the site of
"Max-Planck-Institut f\"{u}r Radioastronomie" (MPIfR) ne\-ar Bonn,
Germany: \url{http://www.mpifr-bonn.mpg.de/survey.html}. Using this
online Survey Sampler allows users to select a region of the sky and
obtain images and data at different frequencies. The 1420-MHz
Stockert survey \citep{reic86} has resolution 0$^\circ$.59, the
820-MHz Dwingeloo survey \citep{berk72} 1$^\circ$.2, the 408-MHz
all-sky survey \citep{hasl82} 0$^\circ$.85, the 34.5-MHz
Gauribidanur survey \citep{dwar90} 0$^\circ$.7 and 22 MHz
\citep{roge99} 1$^\circ$.7. The corresponding observations are given
at the following rates (measured data) for both $l$ and $b$:
$\frac{1^\circ}{4}$ at 1420 MHz, $\frac{1^\circ}{2}$ at 820 MHz,
$\frac{1^\circ}{3}$ at 408 MHz , $\frac{1^\circ}{5}$ at 34.5 MHz and
$\frac{1^\circ}{4}$ at 22 MHz. The effective sensitivities are about
50 mK T$_b$ (T$_b$ is for an average brightness temperature), 0.20
K, 1.0 K, about 700 K and about 500 K, respectively.

We extracted observed brightness temperatures from this data and
then using programs in C and FORTRAN, obtained results shown in this
paper.

The area of HB 21 SNR is enclosed with brightness temperature
contours. The maps of a region in HB 21 SNR, in new Galactic
coordinates ($l$, $b$), with contours of the brightness temperatures
$T_\mathrm{b}$ are plotted in Figs. \ref{fig01}-\ref{fig05}. The
contour lines correspond to the brightness temperatures: minimum,
maximum and nine contours in between. The space between minimum and
maximum contours defines HB 21 SNR. The Galactic longitude and
latitude intervals for HB 21 SNR are the following: $l$ =
[90$^\circ$, 87.5$^\circ$], $b$ = [3.5$^\circ$, 6$^\circ$]. We used
the same method of calculation as given in \citet{bork07} for
Galactic radio loops I--VI, \citet{bork08} for Loops V and VI,
\citet{bork09} for Monoceros loop and in \citet{bork11} for Cygnus
loop.

The mean temperatures and surface brightnesses of this SNR are
computed using data taken from radio-conti\-nuum surveys at 1420,
820, 408, 34.5 and 22 MHz. The areas over which an average
brightness temperature is determined at each of the five frequencies
are taken to be as similar as possible within the limits of
measurement accuracy. However, some differences between these areas
still remain and we think that the major causes of differing borders
between the five frequencies are measured accuracy and small random
and systematic errors in the data. The surface brightness of SNR
must be above the sensitivity limit of the observations and must be
clearly distinguishable from the Galactic background emission
\citep{gree91}. In order to evaluate brightness temperatures over
the SNR we took into account background radiation (see
\citet{webs74}). Borders enclosing the SNR are defined to separate
the SNR from the background.

We have subtracted the background radiation, in order to derive the
mean brightness temperature of the SNR alone. The areas over which
an average brightness temperature is determined at each of the five
frequencies are taken to be as similar as possible within the limits
of measurement accuracy. $T_\mathrm{min}$ from Table \ref{tab01}
means the lower temperature limit between the background and the
SNR, and $T_\mathrm{max}$ means the upper temperature of the SNR. We
used all measured values inside the corresponding regions of $l$ and
$b$, to calculate the brightness temperature of a SNR including the
background. The mean brightness temperature for the SNR is found by
subtracting the mean value of background brightness temperature from
the mean value of the brightness temperature over the area of the
SNR.

After deriving the mean brightness temperatures $T{_\nu}$, we
derived surface brightnesses $\Sigma_{\nu}$ by:

\begin{equation}
\Sigma_\nu = \left( {2k\nu^2 /c^2 } \right) T_\nu.
\label{equ01}
\end{equation}

\noindent where $k$ is Boltzmann constant and $c$ the speed of
light. Results are given in Table \ref{tab01}.

\begin{table}
\centering \caption{Brightnesses of HB 21 reduced to 1000 MHz, using
spectral index $\alpha$ = 0.45 $\pm$ 0.07.}
\begin{tabular}{c|c}
\hline
Frequency & Brightness reduced at 1000 MHz\\
(MHz) & (10$^{-22}$ W/(m$^2$ Hz Sr)) \\
\hline
1420 & 10.83 $\pm$ 0.64 \\
820 & 9.27 $\pm$ 0.24 \\
408 & 11.01 $\pm$ 0.38 \\
34.5 & 14.80 $\pm$ 3.10 \\
22 & 7.88 $\pm$ 2.08 \\
\hline
\end{tabular}
\label{tab02}
\end{table}

\section{Discussion and results}

The results for temperature limits are given in Table \ref{tab01}.
$T_\mathrm{min}$, given in the second column of Table \ref{tab01},
is the lower temperature limit, while $T_\mathrm{max}$ is the upper
temperature limit given in the third column of Table \ref{tab01}.
There are some other sources near the SNR and they do not affect the
calculation. These temperature limits enable us to distinguish the
SNR from background and also from external sources. Then we derived
the temperatures and the surface brightnesses using equation (1) for
each frequency. These results are given in the forth and fifth
columns of Table \ref{tab01}, respectively. Brightnesses of HB 21
reduced to 1000 MHz, using spectral index $\alpha$ = 0.45 $\pm$ 0.07
are given in Table \ref{tab02}.

The radio map of HB 21 SNR at 1420 MHz with contours of brightness
temperature is given in Fig. 1. This SNR has position: $l$ =
[90.$^\circ$, 87.5$^\circ$]; $b$ = [3.5$^\circ$, 6$^\circ$], in new
Galactic coordinates ($l$, $b$). Eleven contours plotted represent
the temperatures $T_\mathrm{min}$ and $T_\mathrm{max}$ from Table
\ref{tab01} and nine contours in between. The contour interval is
0.3 K $T_\mathrm{b}$, starting from the lowest temperature of 6 K up
to 9 K. Also, the corresponding temperature scale is given. In Fig.
2 is presented the radio map of HB 21 SNR at 820 MHz with contours
of brightness temperature. The contour interval is 9.5 K
$T_\mathrm{b}$, starting from the lowest temperature of 16.5 K up to
26 K. The radio map of HB 21 SNR at 408 MHz with contours of
brightness temperature is given in Fig. 3. The contour interval is
5.3 K $T_\mathrm{b}$, starting from the lowest temperature of 87 K
up to 140 K. In Fig. 4 is presented the radio map of HB 21 SNR at
34.5 MHz with contours of brightness temperature. The contour
interval is 3200 K $T_\mathrm{b}$, starting from the lowest
temperature of 49000 K up to 81000 K. The radio map of HB 21 SNR at
22 MHz with contours of brightness temperature is given in Fig. 5.
The contour interval is 3900 K $T_\mathrm{b}$, starting from the
lowest temperature of 105000 K up to 144000 K. The corresponding
temperature scale is given for brightness temperatures of HB 21 at
1420, 820, 408, 34.5 and 22 MHz at Figs.
(\ref{fig01})--(\ref{fig05}), respectively.

\subsection{Spectrum}

The dominant emission mechanism from Galactic supernova remnants
(SNRs) at radio frequencies is synchrotron emission. However in some
SNRs, observations over a very broad range of radio frequencies
reveal a curvature in the spectra of these sources
\citep{leah06,tian05}. The detection of thermal emission at radio
frequencies from Galactic SNRs shows us whether these sources are
interacting with adjacent molecular clouds. It might also be useful
for estimating the ambient density near SNRs using radio continuum
data \citep{uros07}. The presence of a significant amount of thermal
emission will produce a curvature in the observed radio spectrum of
the SNR, particularly for frequencies of 1 GHz and greater
\citep{uros07}.

The radio continuum spectrum of HB 21 SNR (temperature vs
frequency), is presented in Fig. \ref{fig06}. Spectra for five
measurements -- at 22, 34.5, 408, 820 and 1420 MHz are given by
solid line, spectrum obtained from the three frequencies -- 408, 820
and 1420 MHz is given by dashed line, and spectrum obtained from the
two lowest frequencies -- 22 and 34.5 MHz is given by dash-dot line.
The spectrum was generated using mean temperatures at five different
frequencies (see Fig. \ref{fig06}). The brightness temperature
spectral index is defined by Equation (\ref{equ02}). All five
frequencies 1420, 820, 408, 34.5 and 22 MHz, from linear fit, give
$\alpha_5$ = 0.45 $\pm$ 0.07. Frequencies 34.5 and 22 MHz, lie on
very low ends of the spectrum, as presented in Fig. \ref{fig06} and
we analyze them also separately. Derived value $\alpha_2$ = -1.01
(from two frequencies), suggests possible thermal absorption at very
low end of the spectrum.

The curved spectrum of the third order of HB 21, obtained from the
five frequencies is presented in Fig. \ref{fig07}. Obtained spectra
show that a curvature \citep{baar65,milo95} is present in the radio
spectrum at low frequencies, see Figure \ref{fig07}. In the papers
\citet{baar65,milo95} spectra are represented by second order
polynomial fit. For our spectra we used polynomial fit of third
order. This third order polynomial regression line fits properly
temperature versus frequency at five measurements -- at 22, 34.5,
408, 820 and 1420 MHz. The parameters of the third order polynomial
fit are: $a$ = -0.582641 $\pm$ 0.1809 (31.04$\%$), $b$ = 9.38084
$\pm$ 0.2556 (2.725$\%$), $c$ = -5.15063 $\pm$ 0.1133 (2.201$\%$),
$d$ = 0.714124 $\pm$ 0.01609 (2.254$\%$).

We are using two equations:

\begin{equation}
T = K \nu^{-\beta} = K \nu^{-(\alpha + 2)},
\label{equ02}
\end{equation}

\begin{equation}
\log T = a + b \log \nu + c \log^2 \nu + d \log^3 \nu.
\label{equ03}
\end{equation}

Equation (\ref{equ02}) represents the dependence between brightness
temperature, frequency as well as brightness temperature spectral
index $\beta$, ($\beta = \alpha + 2$, where $\alpha$ represents flux
density spectral index) and $K$ is a constant. Equation
(\ref{equ03}) represents third order polynomial fit and tells us
about dependance $T$ of $\nu$. Using equations (\ref{equ02}) and
(\ref{equ03}) and after taking a $log$ of both sides of equations
and derivation of both sides with respect of $\nu$ we obtained
expression:

\begin{equation}
\alpha^c(\nu) = - (b + 2c \log \nu + 3d \log^2 \nu) - 2.
\label{equ04}
\end{equation}

\noindent where index $c$ denotes curved spectrum.

We want to stress that the same procedure can be used if spectra is
given in flux density because relation (\ref{equ01}) holds between
brightness $\Sigma_\nu$ and brightness temperature $T_\nu$ and
between flux density $S_\nu$ and brightness $\Sigma_\nu$ relation
holds:

\begin{equation}
S_\nu = \Sigma_\nu\ \Omega
\label{equ05}
\end{equation}

\noindent where $\Omega$ is the solid angle.

The new equation will be the same as equation (\ref{equ03}), but
with different coefficients.

\begin{equation}
\log S_\nu = a' + b' \log \nu + c' \log^2 \nu + d' \log^3 \nu.
\label{equ06}
\end{equation}

We choose spectrum given in $T_\mathrm{b}$ because our original data
are given in $T_\mathrm{b}$, not in the flux density.

Using these parameters and equation (\ref{equ04}) we can derive the
dependence of spectral index with frequency along the entire
frequency range from 22 to 1420 MHz. Variations of spectral index
$\alpha^c(\nu)$ with frequency in the case of HB 21 is presented in
Fig. \ref{fig08}. Spectral index $\alpha^c(\nu)$ is obtained from
the third order polynomial fit presented in Fig. \ref{fig07}. Open
circles represent values of spectral indices at the five observed
frequencies. We noticed flatter spectral indices at low frequencies
below about 200 MHz. Probably this is due to the thermal mechanism
at frequencies below about 200 MHz or below observed frequency of
408 MHz. Absorption by thermal plasma is probably the preferred
mechanism in the low frequency range of the spectra as it is
proposed by Leahy \citep{leah06}.

From the curved spectrum, we can also calculate spectral index
$\alpha^c$ using equation (\ref{equ04}). In this case $\alpha$ is
function of $\nu$, and therefore for mean $\alpha$ we use:

\begin{equation}
\bar \alpha^c = \frac{1}{\nu_{max} - \nu_{min}}
\int\limits_{\nu_{min}}^{\nu_{max}} {\alpha^c(\nu) d\nu},
\label{equ07}
\end{equation}

\noindent where $\nu_{min}$ = 22 MHz and $\nu_{max}$ = 1420 MHz.

\noindent Using eqs. (\ref{equ04}) and (\ref{equ07}) we obtain $\bar
\alpha^c_{22-1420}$ = 0.46. When we put $\nu_{min}$ = 22 MHz and
$\nu_{max}$ = 34.5 MHz the result is $\bar \alpha^c_{22-34.5}$ =
-0.97.

Since we deal with mean brightness temperature, we lose the
information about index variation because for each frequency we have
only one average value for brightness temperature. Using equation
(\ref{equ02}) we get a connection between brightness temperature and
spectral index $\alpha$. We plot the dependence $\log T$ vs
$\log\nu$. The dependence $\log\nu$ between 408, 820 and 1420 MHz
can be nicely represented by straight line. \citet{leah06} obtained
mean spectral index $\alpha$ = 0.45 using observations presented at
408 and 1420 MHz, and from our Fig. 8 using only points at 408 and
1420 MHz we obtained the $\alpha$ = 0.47 which is in excellent
agreement with Leahy's value. But if we take into account low
frequencies 22 and 34.5 MHz we can find that now dependence $\log T$
vs $\log \nu$ is nicely represented by the polynomial fit of third
order.

This indicates that the spectrum of HB 21 is a combination of
synchrotron and a thermal components. The presence of the additional
component in the radio spectrum of HB 21 suggests that HB 21 is
probably interacting with an adjacent molecular cloud.

\subsection{$T-T$ plot}

The measured data have different resolutions for different
frequencies. In order to obtain $T-T$ plots the data are
retabulated. We convolved data at 1420, 34.5 and 22 MHz to
$0^\circ.25 \times 0^\circ.25$ resolution. Sampling rates of the 820
MHz and 408 MHz survey are much more crude and we did not take it
into account. Then, for each frequency pair we used only the common
points (with the same $(l,b)$) which belong to the loop area at both
frequencies. In that way we reduced loop area to the same area for
different frequencies. The obtained $T-T$ plots for three pairs of
frequencies enabled calculating the spectral indices. We calculated
two $\alpha$ values for each of these tree frequency pairs: between
1420--34.5, 1420--22, 34.5--22 MHz. For each of the tree frequency
pairs, by interchanging the dependent and independent variables we
have obtained two $\alpha$ values for each pair and the mean value
of these fit results is adopted as the radio spectral index, as
suggested in \citet{uyan04}. Regarding two pairs of frequencies
1420--34.5, 1420--22 the average value of spectral index from $T-T$
is $<\alpha_{TT}>_{1420-22,1420-34.5}$ = 0.47 $\pm$ 0.29. Regarding
the lowest pair of frequencies 34.5--22, the average value of
spectral index from $T-T$ is $<\alpha_{TT}>_{34.5-22}$ =
-0.82$\pm$0.82.

It can be noticed that this value agrees well with the corresponding
value obtained from spectrum, as expected (see \citet{uyan04}).

\citet{leah06} determined in HB 21 SNR a mean spectral index of
$0.45$. Our mean spectral index is in agreement with Leahy findings.
He found significant spectral index variations and concluded that
thermal absorption was the preferred mechanism \citep{leah06}. He
concluded that if thermal plasma absorption or ionization losses are
the correct mechanism, at frequencies below 408 MHz should show
flatter spectral indices and if one should be able to distinguish
which mechanism is working, since ionization losses will have no
indices flatter than $0.25$ \citep{leah06}. We obtained flatter
spectral indices then $0.25$ for the HB 21 SNR in the low frequency
domain, and we can expect that thermal plasma absorption is
responsible for the spectral flattening at the lowest radio
frequencies.

\section{Conclusions}

We use observations of the continuum radio emission at 1420, 820,
408, 34.5 and 22 MHz for estimations of the mean brightness
temperatures and surface brightnesses of the HB 21 SNR. The
sensitivity of the brightness temperatures are: 50 mK for 1420 MHz,
0.2 K for 820 MHz, 1.0 K for 408 MHz, about 700 K $T_\mathrm{b}$ for
34.5 MHz and about 500 K $T_\mathrm{b}$ for 22 MHz.

We present the radio continuum spectrum of the HB 21 SNR using
average brightness temperatures at five frequencies. As it can be
seen from Fig. \ref{fig06}, given linear fit provides reliable
spectral index. Our analysis indicates that significant spectral
variations for different frequencies in spectral index are found in
HB 21. We noticed flatter spectral indices at frequencies below 408
MHz as it is proposed by Leahy \citep{leah06}.

This indicates that the spectrum of HB 21 is a combination of
synchrotron and thermal components. The presence of the additional
component in the radio spectrum of HB 21 suggests that this SNR is
interacting with an adjacent molecular cloud.

In \citet{leah06}, observations of the SNR HB 21 are presented at
408 and 1420 MHz. He considered different physical mechanisms for
spectral index variations. He concluded that if the spectra below
408 MHz show flatter spectral indices then absorption by thermal
plasma is the preferred mechanism. Our obtained spectra at 5
frequencies leads to a similar conclusion.

We note an obvious flattening of the spectral indices. In future
work one might investigate a broader frequency range. Probably,
thermal emission at higher frequencies will produce a curvature in
the radio spectrum of the HB 21 SNR, particularly for frequencies
higher then 1 GHz \citep{uros07}.

\begin{acknowledgments}
This research is part of the projects 176003 ''Gravitation and the
large scale structure of the Universe'' and 176005 ''Emission
nebulae: structure and evolution'' supported by the Ministry of
Education and Science of the Republic of Serbia. Authors would like
to thank professor Jack Sulentic for improving English of the paper.
\end{acknowledgments}

\end{document}